\documentclass[aps,prl,reprint,twocolumn,floatfix,showpacs,groupedaddress]{revtex4}
\usepackage{graphicx}
\usepackage{psfrag}
\usepackage{color}
\usepackage{amsbsy}
\usepackage{amssymb}

\usepackage{lineno}

\def\dd{\mbox{d}}

\def\ve{\varepsilon}

\def\r{\rho}

\def\oc{\omega_{\rm c}}
\def\kon{k_{\rm on}}
\def\koff{k_{\rm off}}
\def\kt{k_{\rm t}}

\def\r{{\bf r}}

\begin{document}

\bibliographystyle{biophysj}


\title{Enhancement of charged macromolecule capture
by nanopores in a salt gradient}


\author{Tom Chou}
\affiliation{Dept. of Biomathematics and Dept. of Mathematics, UCLA \\
Los Angeles, CA, 90095-1766, USA}

\begin{abstract}
Nanopores spanning synthetic membranes have been used as key
components in proof-of-principle nanofluidic applications,
particularly those involving manipulation of biomolecules or
sequencing of DNA. The only practical way of manipulating charged
macromolecules near nanopores is through a voltage difference applied
across the nanopore-spanning membrane.  However, recent experiments
have shown that salt concentration gradients applied across nanopores
can also dramatically enhance charged particle capture from a low
concentration reservoir of charged molecules at one end of the
nanopore. This puzzling effect has hitherto eluded a physically
consistent theoretical explanation. Here, we propose an electrokinetic
mechanism of this enhanced capture that relies on the electrostatic
potential near the pore mouth.  For long pores with diameter much
greater than the local screening length, we obtain accurate analytic
expressions showing how salt gradients control the local conductivity
which can lead to increased local electrostatic potentials and charged
analyte capture rates. We also find that the attractive electrostatic
potential may be balanced by an outward, repulsive electroosmotic flow
(EOF) that can in certain cases conspire with the salt gradient to
further {\it enhance} the analyte capture rate.
\end{abstract}

\maketitle


\section{Introduction} 

Recent interest in electrokinetic manipulation of charged
macromolecules has been motivated by technological applications,
particularly those involving sorting and sequencing of nucleic
acids. In a typical realization of single molecule DNA sequencing, an
ionic solution is separated by a membrane with a small pore across the
membrane, connecting two otherwise separated bulk reservoirs
(cf. Fig. \ref{FIG1}). When an electric potential is applied across
the membrane, ionic current flowing through the pore is detected.  DNA
and protein molecules placed on one side of the membrane (the right
reservoir in Fig. \ref{FIG1}), even at low concentrations, can
occasionally block the pore, reducing the ionic current. A time trace
of the ionic current flowing across the pore therefore directly
measures the statistics of blocking and unblocking events.

Experimentally, numerous modifications of the basic configuration
have been studied. 
Biological pores such as $\alpha$-hemolysin have also been chemically
modified to alter internal charges, leading to possible enhancements
of the capture frequency and translocation rates of biopolymers
through the pore \cite{BAYLEY2008}. A number of groups have also
recently fabricated synthetic pores
\cite{GOLOVCHENKO,MARZIALI,DEKKER}, typically through SiN membranes,
for use in macromolecule capture experiments.  Besides pore design,
other approaches to better control macromolecular analyte (both
charged and uncharged) capture and translocation have been explored.
In recent measurements using synthetic pores, an enhanced capture rate
of DNA was observed in the presence of a salt gradient
\cite{MELLERABS,WANUNUABS}.  Salt added to the opposing, non-analyte
reservoir increased the capture rate as a function of the salt ratio
between the two reservoirs.



Much of the theoretical effort, including molecular simulations, has
concentrated on the physics of polymer translocation through the pore
\cite{KLAFTER,PARK,SCHULTEN}.  However, since macromolecular capture
is sensitive to applied voltage, occurs even with uncharged molecules,
and is a stochastic process, the relevant mechanism will involve the
interplay among electrostatics, fluid flow, and the statistics of
capture. Although the electric potentials and flows inside a nanopore
have been presented in the context of Poisson-Nernst-Planck models
\cite{EISENBERG1997,EISENBERG}, the dynamics of initial particle
capture requires a more detailed analysis of the field configuration
in the bulk reservoir, near the pore opening. Finally, the analyte
density in the reservoir needs to be determined \cite{GOLOVCHENKO} in
order to solve the problem of capture of a charged macromolecule to
the pore mouth in the presence of fluid flow and electrokinetic forces
that stochastically switch according to pore blockage. The polymer
capture problem has been treated by Wong and Muthukumar
\cite{WONGMUTHU}, who derive capture rates in the presence of
electroosmotic flow (EOF) induced by an applied voltage bias and
surface charges on the inner pore surface. In this study, as in many
others, \cite{WONGMUTHU,SCHULTEN,EISENBERG1992} the electric field in
the reservoir was neglected and the capture problem was solved only in
the fast translocation limit.

In this paper, we show that bulk electric fields and a more detailed calculation
of the capture problem are necessary to explain the recently-observed
salt gradient-induced enhancement of charged analyte capture rates. By
using geometrical approximations in the high salt limit, we find
analytic expressions for the local salt concentration and
electrostatic potential that are accurate for a range of parameters
relevant to typical experiments. In the presence of an imposed salt
concentration gradient, we show that although most of the potential
drop occurs across thin pores, the electrostatic potential at the pore
mouth plays an important role in the analyte capture rate and cannot
be neglected.  We also show how an EOF {\it into} the analyte
reservoir (by itself convecting macromolecules away from the pore) 
can interact nonlinearly with the salt gradient and 
electrostatic potential to actually {\it increase} the capture rate.
Finally, we analyze implicit solutions for mean capture
rates using the full steady-state Debye-Smulochowski capture problem
with a partially absorbing boundary for the analyte concentration at
the pore mouth.

\section{Electrokinetic equations}

\begin{figure}[htb]
    \begin{center} 
   \includegraphics[width=3.3in]{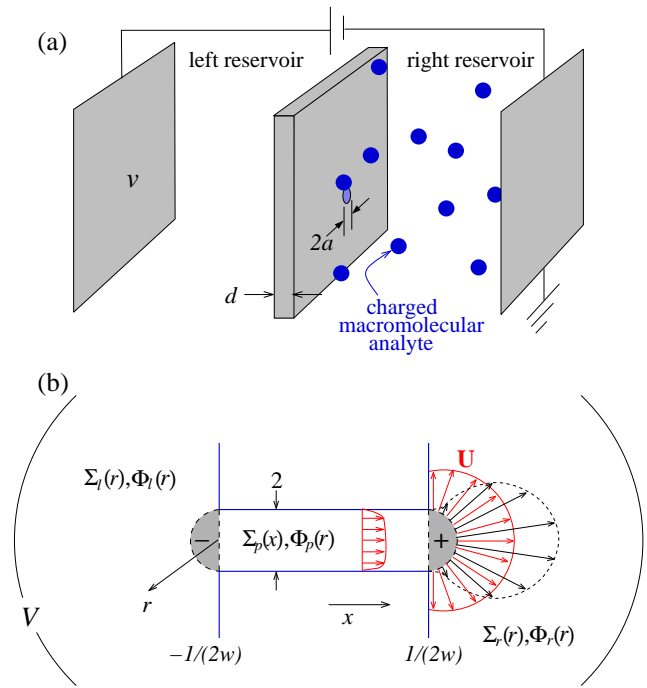}
    \end{center} 
\caption{Schematic of electrokinetic focussing experiments. (a)
Charged analytes are placed in the right reservoir and a voltage bias
is applied. (b) The structure of the pore, electrostatic potential,
and electroosmotically driven fluid flow in dimensionless units, with
distance measured in units of the pore radius. For long pores ($a/d =
w \ll 1$), the concentration and potential fields are approximated as
constant within the small hemispherical regions capping the pore
(denoted by $\pm$). Field and flux continuity conditions are applied
at the hemispherical surfaces. The electroosmotic flow field in the
right chamber will be approximately spherically symmetric (red) if the
membrane surface is uniformly charged but will be more lobe-like
(black) if the membrane flange is uncharged and no-slip boundary
conditions are imposed \cite{WEINBAUM,PARMET}.}
\label{FIG1}
\end{figure}

A typical experimental set-up is depicted in Fig. \ref{FIG1}(a) where
two reservoirs containing aqueous solution are separated by an
electrically insulating membrane containing a single conducting
nanopore through which fluid can flow. A voltage bias $v$ is applied
across electrodes placed far from the pore, resulting in an ionic
current through it.
The full steady-state electrokinetic equations for
the local electrostatic potential $\Phi(\r,t)$, ion concentration
$C_{i}(\r,t)$, and local fluid velocity ${\bf U}(\r,t)$:

\begin{equation}
\nabla\cdot(\epsilon \nabla \varphi(\r,t)) = -4\pi e \sum_{i}z_{i}c_{i}(\r)
\label{POISSON0}
\end{equation}

\begin{equation}
{\bf u}(r)\cdot\nabla c_{i}
= \nabla\cdot(D_{i}\nabla c_{i}) + 
ez_{i}{D_{i}\over k_{B}T}\nabla\cdot(c_{i}\nabla\varphi).
\label{CHARGE0}
\end{equation}

\noindent and 

\begin{equation}
0= -\nabla p +  
\eta\nabla^{2}{\bf u} + e \nabla\varphi \sum_{i}z_{i}c_{i}(\r),
\label{STOKES0}
\end{equation}

\noindent where $z_{i}$ is the valency of solute species $i$,
%
$\epsilon$ is the dielectric permittivity of the solution, $D_{i}$ is the
diffusivity of species $i$, $\eta$ is the dynamic viscosity of the
solution, and $p(\r)$ is the local hydrostatic pressure in the fluid.

For simplicity, we henceforth consider a two component ionic solution
with $z_{\pm}=\pm 1$ where both ion species have the same diffusivity
$D_{+}=D_{-}=D_{\rm{ion}}$.  Reactions at the electrodes will also be
symmetric such that no net charge is built up in the bulk solution. We
analyze Eqs. \ref{POISSON0}-\ref{STOKES0} in the geometry shown in
Fig. \ref{FIG1} where, for mathematical convenience, the electrodes
far from the nanopore are assumed to be hemispherical caps with radius
$r \rightarrow \infty$.

The boundary conditions far away from the nanopore at these
``infinite'' electrodes are $c_{+} = c_{-} = c_{L},\, {\bf u}=0$, and
$\varphi = v$ in the left reservoir, and $c_{+} = c_{-} = c_{R},\,
{\bf u}=0$, and $\varphi = 0$ in the right reservoir.  To obtain
effective boundary conditions at the membrane and on the inner surface of
the pore, we first extract the appropriate physical limit by
defining distance in units of the pore diameter $a$, and
nondimensionalizing Eqs. \ref{POISSON0}-\ref{STOKES0} according to
\begin{equation}
\begin{array}{lll}
\displaystyle C_{\pm} = {c_{\pm}\over c_{R}}, &  
\displaystyle \Phi = {e\varphi \over k_{B}T} & \displaystyle V = {ev \over k_{B}T} \\[13pt]
\displaystyle \displaystyle {\bf U} = {a \over D_{\rm{ion}}}{\bf u}, & \displaystyle P
= {p \over k_{B}T c_{R}} & \displaystyle 
\mu = {D_{\rm{ion}}\eta \over k_{B}T c_{R} a^{2}}.
\end{array}
\end{equation}
Equation ~\ref{POISSON0}, the sum and difference of the $z=+1$ and $z=-1$ components
of Eq. \ref{CHARGE0}, and Eq.~\ref{STOKES0} become, respectively,


\begin{eqnarray}
\displaystyle \nabla\cdot(\epsilon\nabla \Phi(\r)) + \Lambda_{R} Q(\r) = 0 \label{SSPOISSONQ}\\
\displaystyle \nabla\cdot\left[\nabla\Sigma(\r) + Q(\r)\nabla\Phi(\r) - {\bf U}(\r)\Sigma(\r)\right]=0 \label{SSIONSUM} \\
\displaystyle \nabla\cdot\left[\nabla Q(\r) + \Sigma(\r)\nabla\Phi(\r) - {\bf U}(\r)Q(\r)\right] = 0 \label{SSIONDIFF} \\
\displaystyle \mu\nabla^{2}{\bf U}(\r) - \nabla P(\r) + Q(\r)\nabla\Phi(\r) = 0. \label{SSSTOKESQ}
\end{eqnarray}

\noindent Above, we define  the ion concentration difference 
$Q(\r) = {1\over 2}(C_{+}(\r) - C_{-}(\r))$, sum
$\Sigma(\r)={1\over 2}(C_{+}(\r)+C_{-}(\r))$, and

\begin{equation}
\Lambda_{R} \equiv {8\pi e^{2} c_{R}a^{2} \over k_{B}T} \equiv (\kappa_{R} a)^{2}.
\end{equation}

\noindent The quantity $\kappa_{R}$ represents the inverse ionic
screening length associated with the ionic solution deep in the right
reservoir.  For sufficiently high salt concentrations $c_{R} \sim
1-0.1$M, the corresponding screening length $\kappa_{R}^{-1} \sim
0.3-1$nm, while synthetic pores have radii on order of at least
$a=3-5$nm, rendering $\Lambda \gtrsim 10$ large. At distances of least
a screening length away from the membrane or pore surfaces, this
separation of scales allows us to consider only the ``outer''
solutions associated with charge-neutral conducting fluid bulk
reservoirs where boundary layers of charge separation do not reach.

In the $\Lambda_{R}^{-1} \equiv \ve \rightarrow 0$ limit,
Eq. \ref{SSPOISSONQ} represents a singular perturbation.  The outer
solutions to the system of equations can be found
by considering expansions of the form 

\begin{equation}
\begin{array}{rcl}
\Phi(\r) & = & \Phi_{0}(\r)+ \ve \Phi_{1}(\r) + \ve^{2}\Phi_{2}(\r) + \ldots \nonumber \\
\Sigma(\r) & = &  \Sigma_{0}(\r)  + \ve \Sigma_{1}(\r) + \ve^{2}\Sigma_{2}(\r) + \ldots \nonumber \\
Q(\r) & = & Q_{0}(\r)  + \ve Q_{1}(\r)  + \ve^{2} Q_{2}(\r)  + \ldots \nonumber \\
{\bf U}(\r) & = & {\bf U}_{0}(\r)  + \ve {\bf U}_{1}(\r)  + \ve^{2}{\bf U}_{2}(\r)  + \ldots \nonumber  \\
P(\r) & = & P_{0}(\r)  + \ve P_{1}(\r)  + \ve^{2}P_{2}(\r)  + \ldots.
\end{array}
\end{equation}
%
Upon using these expansions in Poisson's equation,
$\ve\nabla\cdot(\epsilon \nabla\Phi) = Q$, we find an outer solution
$Q_{0} = 0$, as expected in the charge-neutral limit of a fluid
conductor. To find solutions accurate to $O(\ve^{0})$, we must solve
the remaining equations

\begin{eqnarray}
\displaystyle \nabla\cdot\left[\nabla\Sigma_{0}(\r)-{\bf U}_{0}(\r)\Sigma_{0}(\r)\right]=0 \label{SIGMA0}\\
\displaystyle \nabla\cdot\left[\Sigma_{0}(\r)\nabla\Phi_{0}(\r)\right] = 0 \label{CONDUCTION}\\
\mu\nabla^{2}{\bf U}_{0}(\r) - \nabla P_{0}(\r) = 0 \label{STOKES}
\end{eqnarray}
%
Henceforth, we consider only the zeroth order solutions and drop the
subscript $(0)$ notation.  Equation \ref{SIGMA0} simply describes the
steady-state convection-diffusion of the total local salt concentration
$\Sigma(\r)$.  The ionic conductivity of a locally neutral electrolyte
solution is proportional to this local salt concentration.
Eq. \ref{CONDUCTION} is an expression of Ohm's law that includes a
spatially varying ionic conductivity.

In the relevant limit of small screening length, the near-field
boundary conditions for the concentration and potentials associated
with the outer Eqs. \ref{SIGMA0}-\ref{CONDUCTION} can be obtained by
noting that the inner solution near the membrane decays exponentially
({\it e.g.}, as does the solution to the Poisson-Boltzmann equation).
To match these exponentially decaying inner solutions requires
effective Neumann boundary conditions for the outer solutions of
$\Sigma$ and $\Phi$ near the solution-membrane interface. Such
boundary conditions embody the impenetrable nature of the membrane to
both the salt and the ionic current, and has been more formally
derived in \cite{EISENBERG}.  The Neumann boundary conditions allow
the use of spherical symmetry of the solutions within each reservoir,
rendering Eqs. \ref{SIGMA0} and \ref{CONDUCTION} one-dimensional in
the radial coordinate.  The far-field boundary conditions on the
dimensionless quantities are $\Sigma_{r}(r\rightarrow \infty) = 1$,
$\Sigma_{\ell}(r\rightarrow \infty) \equiv \Sigma_{L}$, and
$\Phi_{\ell}(r\rightarrow \infty) = V$, $\Phi_{r}(r\rightarrow \infty)
= 0$.  Although complicated expressions have been derived to compute
the exact concentration profiles associated with steady-state
diffusion through a finite width pore \cite{KELMAN}, our spherical
approximation dramatically simplifies the analysis of
Eqs. \ref{SIGMA0}-\ref{CONDUCTION}.

Finally, we must also consider the structure of the flow field ${\bf
U}$ in Eq. \ref{STOKES}, even if no hydrostatic pressure gradient is
externally applied and $P(r\rightarrow \infty) = 0$. A nonzero outer
velocity field will arise from boundary charge-induced electroosmosis,
in which a surface charge on the inner surface of the pore enhances
the concentration of the counterions near the inner surface the pore.
An applied electric then pushes this slightly charged layer, dragging
the bulk fluid with it.  Matching the outer velocity field with the
electric field-driven, inner layer of fluid within a Debye screening
length of the charged inner pore surface results in a plug-like outer
flow velocity profile. The plug-like profile allows us to consider the
outer solution for ${\bf U}$ inside the pore as a constant
$U\hat{x}$. The magnitude $U$ of the plug flow velocity is
proportional to the ``$\zeta$-potential'' at the pore surface and the
electric field applied across the pore \cite{EOF}.  Within
linearized electrostatics described by the Debye-H\"{u}ckel equation,
the $\zeta$-potential is proportional to the surface charge density
and the local screening length of the solution.

In the absence of pondermotive body forces on the fluid outside the
pore, the velocity field ${\bf U}$, under no-slip boundary conditions,
has been calculated as a series expansion \cite{WEINBAUM} and in terms
of integrals over Bessel's functions \cite{PARMET}.  The lobe-like
flow patterns (cf. Fig. \ref{FIG1}(b)) were obtained using no-slip
boundary conditions at the membrane interface and are not spherically
symmetric.  However, if the membrane walls are also uniformly charged,
potential gradients along the wall exterior to the pore will drive an
inner EOF along the wall, giving rise to an effective slip boundary
condition for the outer flow field ${\bf U}$ near the wall. We will
show that the outer-solution electric field tangential along the
membrane falls off as $1/r^{2}$, plus logarithmic corrections in the
presence of salt gradients. Therefore, the velocity profile in the
reservoirs can be self-consistently approximated by a radial profile
${\bf U} = U\hat{r}/r^{2}$ obeying incompressibility. The actual
velocity profile will qualitatively resemble radial flow, while
retaining some features of a lobed-flow profile. Since we expect our
gross results will be sensitive mostly to the typical magnitude of
${\bf U}$, and not to the details of its angular dependence, we will
also assume a simple radial form for ${\bf U}(\r)$.

Since we assume radial symmetry in the outer solutions of all
quantities, we apply uniform boundary conditions at both the near and
far boundaries at $r \rightarrow \infty$ and $r=1$, respectively. Upon
defining the variables

\begin{equation}
G \equiv \exp\left[{U\over 2}\right]  \quad \mbox{and}\quad 
H \equiv \exp\left[{U d \over a}\right] \equiv\exp\left[{U \over w}\right],
\label{GH}
\end{equation}
where $w \equiv a/d \ll 1$ is the pore aspect ratio, the total ion
concentration can be found in terms of the normalized bulk salt
concentration $\Sigma_{L}$ far from the pore in the left reservoir.
When the pore is unblocked by analyte, salt freely diffuses
across. Using the general solutions of Eq. \ref{SIGMA0} in each region
(left and right reservoirs, and pore), and imposing conservation of
ion flux at the surfaces of the hemispherical caps between these
regions (cf. Fig. \ref{FIG1}(b)), $2\pi \partial_{r}\Sigma_{\ell}(r=1)
+ \pi \partial_{x}\Sigma_{p}(x=-1/(2w)) = 0$ and $2\pi
\partial_{r}\Sigma_{r}(r=1) - \pi \partial_{x}\Sigma_{p}(x=1/(2w))=
0$, we find

\begin{equation}
\displaystyle \Sigma_{\ell}(r) = {\Sigma_{L}GH -1\over GH-1} 
-{(\Sigma_{L}-1)\over GH-1}\exp\left[{U \over 2r}\right],
\label{SIGMAl}
\end{equation}

\begin{equation}
\displaystyle \Sigma_{p}(x) = {\Sigma_{L}GH -1\over GH-1} 
- {(\Sigma_{L}-1)\sqrt{GH}\over GH-1}e^{Ux},
\label{SIGMAp}
\end{equation}
and

\begin{equation}
\displaystyle \Sigma_{r}(r) = {\Sigma_{L}GH -1\over GH-1}
-{(\Sigma_{L}-1)GH\over GH-1}\exp\left[-{U \over 2r}\right].
\label{SIGMAr}
\end{equation}
Although exact solutions can be computed
\cite{KELMAN}, for small $w\ll 1$, the simple forms given in
Eqs. (\ref{SIGMAl}-\ref{SIGMAr}) are accurate to $O(w)$.

The salt concentration $\Sigma(\r)$ determines the local ionic
conductivity as a function of the EOF velocity $U$ and the
experimentally imposed bulk salt ratio $\Sigma_{L}$. We now substitute
the approximate solutions for $\Sigma(r)$ into Eq. \ref{CONDUCTION} to
find the electrostatic potential $\Phi(\r)$.  Denoting the ion
concentration and electrostatic potential within the hemispheres
capping the pore as $\Sigma_{p}(1/(2w))=\Sigma_{r}(1)\equiv
\Sigma_{+}$, $\Sigma_{p}(1/(2w))=\Sigma_{\ell}(1)\equiv \Sigma_{-}$,
and $\Phi_{p}(1/(2w))=\Phi_{r}(1)\equiv \Phi_{+}$,
$\Phi_{p}(-1/(2w))=\Phi_{\ell}(1)\equiv \Phi_{-}$, and applying all
far-field boundary conditions, we find

\begin{equation}
\begin{array}{rl}
\Phi_{\ell}(r) & \displaystyle = (V-\Phi_{-}){\int_{1}^{r}(y^2 \Sigma_{\ell}(y))^{-1} \dd y \over
\int_{1}^{\infty}(y^{2}\Sigma_{\ell}(y))^{-1}\dd y} + \Phi_{-} \\[15pt]
\: & \displaystyle =V{\ln\left({\Sigma_{\ell}(r)\over \Sigma_{-}}\right)+{U \over 2}-{U \over 2r} \over
\ln\left({\Sigma_{L}\over \Sigma_{-}}\right) + {U\over 2}}+ 
\Phi_{-}{\ln\left({\Sigma_{L}\over \Sigma_{\ell}(r)}\right) + {U \over 2r} \over 
\ln\left({\Sigma_{L}\over \Sigma_{-}}\right) + {U \over 2}},
\label{PHIL}
\end{array}
\end{equation}
\begin{equation}
\begin{array}{rl}
\Phi_{p}(x) & \displaystyle = (\Phi_{+}-\Phi_{-})
{\int_{-1/(2w)}^{x}\Sigma_{p}^{-1}(y) \dd y \over
\int_{-1/(2w)}^{1/(2w)}\Sigma^{-1}_{p}(y)\dd y} + \Phi_{-} \\[16pt] \:
& \displaystyle = (\Phi_{+}-\Phi_{-}){\ln\left({\Sigma_{-}\over
\Sigma_{p}(x)}\right) + Ux + {U \over 2w} \over \ln\left({\Sigma_{-}\over
\Sigma_{+}}\right) + {U \over w}} +\Phi_{-},
\label{PHIP}
\end{array}
\end{equation}
and
\begin{equation}
\begin{array}{rl}
\Phi_{r}(r) & \displaystyle = \Phi_{+}-\Phi_{+} {\int_{1}^{r}(y^2 \Sigma_{r}(y))^{-1}\dd y \over 
\int_{1}^{\infty}(y^2\Sigma_{r}(y))^{-1}\dd y} \\[15pt]
\: & \displaystyle = \Phi_{+}{\ln \Sigma_{r}(r) + U/(2r) \over \ln \Sigma_{+}+U/2}.
\label{PHIR}
\end{array}
\end{equation}

In order to explicitly determine the potentials $\Phi_{\pm}$, we apply
Kirchhoffs Law,

\begin{equation}
\begin{array}{l}
2\pi \Sigma_{-}\partial_{r}\Phi_{\ell}(r=1)+ \pi \Sigma_{-}\partial_{x}\Phi_{p}(x=-d/(2a)) = 0,\\[13pt]
2\pi \Sigma_{+}\partial_{r}\Phi_{r}(r=1) - \pi \Sigma_{+}\partial_{x}\Phi_{p}(x=d/(2a))= 0
\end{array}
\end{equation}
conserving ionic current  
across the pore mouths to find 

\begin{equation}
\begin{array}{rl}
\displaystyle \Phi_{+} & \displaystyle = V{\ln \Sigma_{+} +U/2 \over
\ln \Sigma_{L} + U (1+1/w)} \\[13pt]
\end{array}
\label{PHIPLUS}
\end{equation}

\noindent and 

\begin{equation}
\begin{array}{rl}
\displaystyle \Phi_{-} = V{\ln\Sigma_{-}  + U(1+2/w)/2 \over 
\ln \Sigma_{L} + U(1+1/w)}
\end{array}
\label{PHIMINUS}
\end{equation}
In the limit of vanishing salt gradient, $\Sigma_{L} \rightarrow 1$,
and

\begin{equation}
\Phi_{+}(\Sigma_{L} \rightarrow 1, w, U) 
\simeq {w V \over 2(1+w)} + O(\Sigma_{L}-1),
\end{equation}
while for vanishing electroosmotic flow, $U \rightarrow 0$ and

\begin{equation}
\Phi_{+}(\Sigma_{L},w,U\rightarrow 0) 
\simeq {V \over \ln \Sigma_{L}}\ln\left[{2+w(\Sigma_{L}+1) \over 
2(w+1)}\right].
\end{equation}
In the case $\Sigma_{L} \rightarrow \infty$, the negligible resistance
offered by the left reservoir makes $\Phi_{+}$ slowly (logarithmically)
approach $V$. 



\section{Results and Analysis}

\subsection{Concentrations and potentials}

\begin{figure}[htb]
    \begin{center} 
     \includegraphics[width=3.4in]{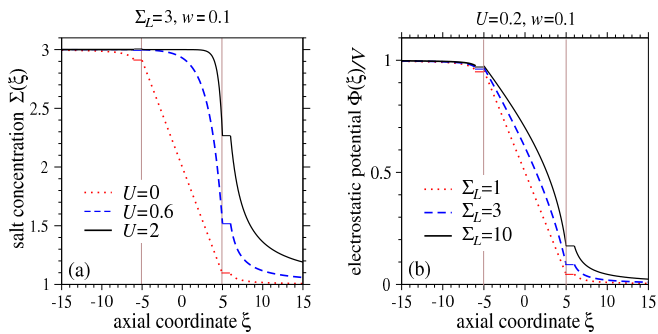}
    \end{center} 
\caption{Salt concentration and electrostatic potential across an
unimpeded pore of aspect ratio $w=a/d=0.1$. (a) The salt density
$\Sigma(\xi)$ as a function of the axial coordinate for various EOF
velocities $U$ and salt ratio $\Sigma_{L}=3$. (b) The normalized
potential $\Phi(\xi)/V$ for various salt ratios $\Sigma_{L}$.}
\label{FIG2}
\end{figure}

Fig. \ref{FIG2}(a) shows the total salt concentration along the axial
coordinate, made up of solutions to $\Sigma_{\ell}(r)$,
$\Sigma_{p}(x)$, and $\Sigma_{r}(r)$.  For zero EOF ($U=0$), the total
salt concentration is linear in the pore region, but has a decaying
structure in the two reservoirs.

%
The spatially varying total salt concentration results in a spatially
varying conductivity. Fig. \ref{FIG2}(b) shows the resulting
electrostatic potential along the axial coordinate $\xi$. For
vanishing salt gradient $\Sigma_{L}=1$, Eq. \ref{PHIR} shows the
potential decays as $1/r$ in the analyte reservoir, consistent with
our assumption of a hemispherical EOF profile far from the pore.  In
the presence of an imposed salt gradient ($\Sigma_{L} \neq 1$),
additional logarithmic terms arise in the reservoir potential
$\Phi_{r}(r)$. These results show that although most of the potential
drop occurs across the membrane-spanning pore, in the presence of
ionic current, the electrostatic potentials in the reservoirs vary
slowly with distance from the pore mouths. 

For $\Sigma_{L}=1$ and long pores ($w \ll 1$), the potential at the
pore mouth in the grounded analyte reservoir $\Phi_{+}\approx wV$ may
not be small for sufficiently large applied bias voltage.
Furthermore, we will show that the analyte capture rate is sensitive
to $\Phi_{+}$ as it depends on the exponent of $\Phi_{+}$.  For
imposed salt gradients $\Sigma_{L}>1$, the conductivity in the left
reservoir is relatively higher and more of the voltage drop occurs
across the pore and the analyte (right) reservoir. This further raises
the potential $\Phi_{r}(r=1) = \Phi_{+}$ felt by the charged
macromolecules at the mouth of the pore in the analyte chamber.  Note
that the potential $\Phi_{r}(r)$ depends on the gradient of
conductivity (arising from the salt concentration gradient), and not
on its absolute value.  The conductance variation arising from salt
concentration gradients provides a simple mechanism by which the
potential $\Phi_{r}(r)$ can be increased through the salt ratio
$\Sigma_{L}$, enhancing in the capture rate of charged analytes.

\subsection{Effect of electroosmotic flows}

Now consider how the details of the electroosmotic flow velocity $U$ may
affect the electrostatic potential. For an EOF with given magnitude
$U$, the potentials $\Phi_{\pm}$ can be calculated from
Eqs. \ref{PHIPLUS} and \ref{PHIMINUS}.  However, recall that the EOF
velocity is driven by the potential difference applied across the pore
\cite{EOF}; therefore, $U$ must be self-consistently solved by finding
the root to

\begin{equation}
U = \Gamma(U,\Sigma_{L}) (\Phi_{+}[U,\Sigma_{L}] - \Phi_{-}[U,\Sigma_{L}]),
\label{EOFEQN}
\end{equation}
where we have explicitly denoted the dependence of $\Phi_{\pm}$ on $U$
and $\Sigma_{L}$. The prefactor $\Gamma(U,\Sigma_{L})$ measures the
effective electroosmotic permeability, which is inversely proportional
to the pore length and fluid viscosity, and proportional to the
``$\zeta$-potential'' \cite{EOF}. Within the linearized
Debye-H\"{u}ckel theory for electrolytes, this local $\zeta$-potential
is proportional to the pore surface charge times the local screening
length $\kappa^{-1}(x)$. In our problem where the ionic strength is
varying in the axial direction along the pore, the $\zeta-$potential
is also varying along the pore.  Although a nonuniform surface
potential, along with the constraint of fluid incompressibility, can
give rise to nonuniform flow within the pore, it has been shown that
the {\it net} fluid flow across a pore can be found by averaging the
$\zeta$-potential (or screening length) across the pore
\cite{IEEE,IDOL,KENNY}:

\begin{equation}
\begin{array}{rl}
\Gamma(U,\Sigma_{L}) & \displaystyle \equiv \left({\sigma k_{B}T \over \eta d e}\right){1\over d}
\int_{-d/2}^{d/2}{\dd x \over \kappa(x)} \\[13pt]
\: & \displaystyle \equiv \Gamma_{R}\, 
w\int_{-1/(2w)}^{1/(2w)}{\dd x\over \sqrt{\Sigma_{p}(x)}},
\label{S}
\end{array}
\end{equation}
where 

\begin{equation}
\Gamma_{R}\equiv \left({\sigma \over e a c_{R}}\right)
{w \over \mu \Lambda_{R}^{1/2}} \nonumber
\end{equation}
is the dimensionless permeability referenced to the salt in the right
chamber. For a typical SiN membrane with constant surface charge
density $\vert \sigma\vert < 1$ microcoulomb/cm$^{2}$
\cite{SINCHARGE}, a 5nm radius, 40nm length pore in water gives $\vert
\Gamma_{R}\vert < 0.2$. 


\begin{figure}[htb]
    \begin{center} 
     \includegraphics[width=3.4in]{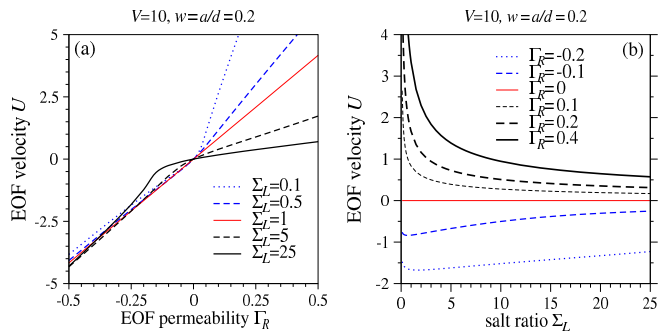}
    \end{center} 
\caption{(a) EOF velocity $U$ as a function of effective pore EOF
permeability $\Gamma_{R}$. The response deviates from linear for large
and small salt ratios. The deviations are most pronounced for
$\Gamma_{R}>0$ where EOF is into the right reservoir ($U>0$). (b) EOF
velocity as a function of salt ratio for various effective pore
surface charge densities. For $\Gamma_{R}>0$, increasing the salt
ratio decreases the effective screening length in the pore, reducing
the EOF velocity (cf. Eq. \ref{EOFEQN}). When $\Gamma_{R}<0$, the salt
in the right reservoir is swept into the left reservoir, keeping the
screening length approximately $\kappa^{-1}_{R}$ throughout the pore
and very little dependence on $\Sigma_{L}$ arises.  In both plots
$V=10$ and $w=a/d=0.2$.}
\label{FIG3}
\end{figure}

Figure \ref{FIG3}(a) shows the self-consistent EOF velocity $U$ 
(obtained by using Eq.~\ref{S} and solving Eq.~\ref{EOFEQN}) as a
function of $\Gamma_{R}$, for various salt ratios $\Sigma_{L}$. For
$\Gamma_{R} > 0$ and $U>0$, salt is being swept from the left to right
reservoirs. When $\Sigma_{L} > 1$, the pore feels a higher averaged
salt concentration, lowering the $\zeta$-potential, thereby reducing
the EOF response to $\Gamma_{R}$. Conversely, when $\Sigma_{L} <1$, a
higher $\zeta$-potential and stronger response arises.
Fig. \ref{FIG3}(b) shows the EOF velocity $U$ as a function of salt
ratio $\Sigma_{L}$ for various $\Gamma_{R}$. For
$\Sigma_{L}\rightarrow \infty$, $U$ vanishes along with the effective
pore $\zeta$-potential.

\begin{figure}[htb]
    \begin{center} 
     \includegraphics[width=3.4in]{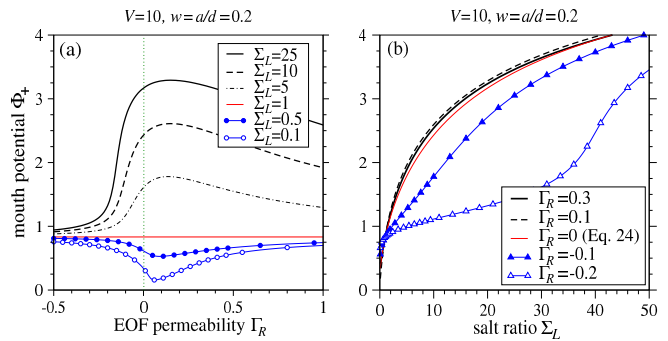}
    \end{center} 
\caption{(a). The electrostatic potential in the right reservoir
(normalized by the applied potential $V$), as a function of salt
asymmetry $\Sigma_{L}$ for various pore EOF permeabilities
$\Gamma_{R}$. Here the pore aspect ratio was set to $w=a/d = 0.2$. (b)
The magnitude of the potential near the mouth pore as a function of
EOF permeability $\Gamma_{R}$ for various salt ratios.}
\label{FIG4}
\end{figure}

The pore mouth potential $\Phi_{+}$ felt by the charged analyte is
shown in Fig. \ref{FIG4}. As a function of pore charge/permeability
$\Gamma_{R}$, the potential $\Phi_{+}$ exhibits a maximum(minimum) for
$\Sigma_{L}>1 (\Sigma_{L}<1)$.  This nonmonotonic dependence arises
because for $\Sigma_{L}>1$ and small $U\gtrsim 0$, the EOF changes the
conductance structure so that $\Phi_{+}$ initially increases.  In
other words, the largest voltage drop across the analyte (right
reservoir) reservoir occurs at small, positive $U$.  For larger $U$,
high salt is swept well into the right reservoir reducing the
effective relative conductivity across the pore. Most of the voltage
drop then occurs in regions away from the pore well in the right
reservoir, diminishing $\Phi_{+}$ to the value $wV/(2+2w)$ expected in
the uniform salt ($\Sigma_{L}=1$) limit. For $\Sigma_{L}<1$, more
voltage drop occurs across the left reservoir, reducing $\Phi_{+}$
below $wV/(2+2w)$.
Fig. \ref{FIG4}(b) shows that as the pore potential $\Phi_{+}$
increases with the salt ratio $\Sigma_{L}$. This is consistent with
the observed increase in analyte capture rate with increasing salt in
the analyte-free chamber. Also consistent with Fig. \ref{FIG4}(a) is
the slight increase, then decrease in $\Phi_{+}$ as $\Gamma_{R}$ is
increased


\subsection{Charged analyte distribution}

We now model how both the approximate EOF velocity $\hat{r}
U/(2r^{2})$ and the electrostatic potentials (Eqs. \ref{PHIL},
\ref{PHIR}, and \ref{PHIP}) affect the capture of charged analytes to
the nanopore as functions of parameters such as the applied salt ratio
$\Sigma_{L}$, applied voltage bias $V$, and and pore surface
charge/permeability $\Gamma_{R}$. When a charged analyte molecule of
size of order $a$ enters the hemispherical cap, it blocks the pore and
prevents ion transport. Such a nonspecifically adsorbed particle can
spontaneously desorb from the mouth of the pore with rate $\koff$.
Alternatively, as in the case of DNA, it may translocate through the
pore to the opposite, receiving reservoir.  Although translocation of
a polymer involves many stochastic degrees of freedom, we will lump
the process into a single, effective rate $\kt$, such that $\kt^{-1}$
represents the typical time for the macromolecule to fully tunnel
across the pore, allowing ionic current to flow again.

If the hemispherical cap can accommodate at most one blocking
macromolecule, its mean field, steady-state occupation $0\leq \theta
\leq 1$ is balanced according to:

\begin{equation}
\kon \rho(1)(1-\theta) = (\koff + \kt)\theta,
\label{THETA}
\end{equation}
where $\rho(1) \equiv \lim_{r\rightarrow 1^{+}}\rho(r)$ is the analyte
concentration just outside the cap region, and $\kon$ is its
adsorption rate into the hemisphere. To explicitly determine $\theta$
we need to relate the unmeasurable, kinetically-determined $\rho(1)$
with the experimentally imposed macromolecular density $\rho_{\infty}
\equiv \rho(r\rightarrow \infty)$. This relationship is obtained by
solving the mean field, steady-state convection-diffusion equation for
the density in the bulk region:

\begin{equation}
(1-\theta)\nabla\cdot\left[{\bf A}(r)\rho(r)\right] = \nabla^{2}\rho(r), \quad r > 1,
\label{RHODIFF}
\end{equation}
where 

\begin{equation}
{\bf A}(r) = A(r)\hat{r} = \left[{U\over 2Dr^{2}} + 
q{\partial \Phi_{r}(r)\over \partial r}\right]\hat{r}
\end{equation}
is the normalized drift arising from both a hydrodynamic flow and an
electrostatic potential induced by electroosmosis and ionic
conduction, respectively. The drift induced by the electrostatic
potential is proportional to the effective number of electron charges
$q$ of the analyte. The factor $(1-\theta)$ reflects our
assumption that the EOF flow and ionic current is completely shut off
when a macromolecule occupies the cap ($\theta=1$). In writing
Eqs. \ref{THETA} and \ref{RHODIFF}, we have implicitly assumed that
the time $\tau_{\rm eff}$ required to form the effective potential
$U/(2Dr) + q\partial_{r}\Phi_{r}(r)$ is much less than the typical
times associated with dissociation, association, or translocation:
$\tau_{\rm eff} \ll \koff^{-1}, \kon^{-1}, \kt^{-1}$. For relevant parameters,
$\tau_{\rm eff} \sim a^{2}/(\Lambda_{R} D) \sim 10^{-10}$s. Provided all
the rates are slower than the rate of reestablishing the effective
potential, we can assume that the field and flow configurations
instantaneously follow those corresponding to whether or not the pore is
blocked.

The boundary condition associated with Eq.~\ref{RHODIFF} is applied
just outside the hemispherical cap ($r=1^{+}$) and is determined by
macromolecular flux balance through the interface
\cite{ADSORPTION1,ADSORPTION2}

\begin{equation}
\begin{array}{l}
D\partial_{r}\rho(r)\big|_{r=1^{+}} - (1-\theta)\left[{U \over 2} + Dq\nabla\Phi_{+}\right]\rho(1)\\[13pt]
\:\hspace{2.7cm} = \kon\rho(1)(1-\theta) - \koff \theta.
\end{array}
\end{equation}
Upon solving Eq. \ref{RHODIFF} for $\rho(r)$ and using Eq. \ref{THETA}
for $\rho(1)$, we find

\begin{widetext}
\begin{equation}
\rho(r) = e^{(1-\theta)\int_{1}^{r} A(y)dy}\left[{\kt \theta \over D}
\int_{1}^{r} y^{-2}e^{-(1-\theta)\int_{1}^{y}A(y')dy'} dy
+ {(\koff+\kt)\theta \over \kon (1-\theta)}\right].
\label{RHOSOLN}
\end{equation}
\end{widetext}
After setting $r\rightarrow \infty$ in Eq. \ref{RHOSOLN}, we relate
the mean cap occupation $\theta$ to the bulk analyte density
$\rho_{\infty}$ through the physical solution of the
integro-transcendental equation

\begin{equation}
{\kt \theta \over D}I(\theta)+ {(\koff + \kt)\theta \over 
\kon (1-\theta)}e^{(1-\theta)(U/(2D)-q\Phi_{+})} = \rho_{\infty},
\label{TRANS}
\end{equation}
where 

\begin{equation}
I(\theta) \equiv \int_{1}^{\infty}e^{(1-\theta)\left({U\over 2Dr} - q\Phi_{r}(r)\right)}r^{-2} \dd r.
\end{equation}

When $\kt=0$, adsorbed macromolecules do not translocate and can only
detach back into the right bulk reservoir.  In this limit, $I(\theta)$
does not arise in Eq.~\ref{TRANS} and $\theta$ depends only on the
value $\Phi_{+}$ of electrostatic potential at the pore mouth.
Moreover, the dependence on the normalized macromolecule diffusivity
arises only in $U/(2D)$, the drift due to EOF at the pore
mouth. Although the EOF and ionic conduction in the fluid arises from
nonequilibrium processes, the density profile $\rho(r)$ in the $k_{\rm
t} = 0$ limit is an {\it equilibrium} density self-consistently
determined by the effective potential $(1-\theta)\left[U/(2Dr) -
q\Phi_{r}(r)\right]$.  For parameters such that $A(1) \equiv
\left[U/(2D)-Q\Phi_{+}\right] \gg \ln(\rho_{\infty}\kon/\koff)$,

\begin{equation}
\theta \sim \left({\kon \over \koff}\right)\rho_{\infty}e^{-U/(2D)+q\Phi_{+}} \ll 1.
\label{THETAEQ}
\end{equation}
This form corresponds to an equilibrium ``adsorption isotherm'' on the
single site and depends only on the value of the potential energy at that
site.





Conversely, in the limit of high translocation rates such that
$\kt\kon/D \gg \koff$ and $\left[U/(2D)-q\Phi_{+}\right] \gg
\ln(\rho_{\infty}\kon/\kt)$, the first iteration about $\theta = 0$ to
Eq. \ref{TRANS} yields

\begin{equation}
\theta \sim {(\kon/\kt) D\rho_{\infty} e^{-U/(2D)+q\Phi_{+}} \over 
\kon I(0) e^{-U/(2D)+q\Phi_{+}} + D}.
\label{THETAABSORB}
\end{equation}
Here, the surface density $\rho(1)\sim \kt\theta/\kon$ resembles that
of an adsorbing sphere with an attachment rate $\kon$, but modified by
the $O(1)$ term $I(0) e^{-U/(2D)+q\Phi_{+}}$ resulting from
stochastically switching of the effective drift.

Although the full solution for the pore occupation must be solved
numerically, we see that $\theta$ is exponentially sensitive to both
the magnitude of the EOF velocity $U$ and the electrostatic potential
at the cap $\Phi_{+}$ through an effective drift defined by the
combination $U/(2D)-q\Phi_{+}$.
Note that both $U$ and $\Phi_{+}$ depend linearly on the bias voltage
$V$, but nonlinearly on the salt ratio $\Sigma_{L}$.  The EOF velocity
$U$ is a function of $\Gamma_{R}$ through the solution
Eq. \ref{EOFEQN}, while $\Phi_{+}$ depends indirectly on $\Gamma_{R}$
through the resulting flow velocity $U$ that changes the local
conductivity when $\Sigma_{L}\neq 1$. However, only the electrostatic
drift $q\Phi_{+}$ depends on the effective analyte charge $q$.

\begin{figure}[htb]
    \begin{center} 
     \includegraphics[width=3.4in]{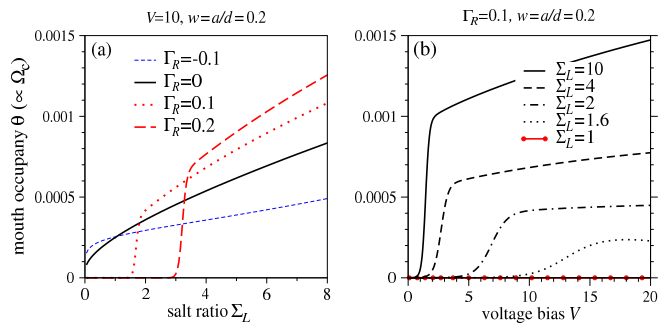}
    \end{center} 
\caption{(a) Pore mouth occupation fraction $\theta$ as a function of
salt ratio $\Sigma_{L}$ for various EOF permeability $\Gamma_{R}$ and
$V=10$. Note the sharp increase in $\theta$ as a function of
$\Sigma_{L}$ for positive surface charges.  Parameters used were
$\rho_{\infty}=10^{-7}$, $V=10$, $w=0.2$, $\kon/\koff=1000$, and
$\kon/\kt=10^{4}$.  For small $\Sigma_{L}$, larger $\Gamma_{R}>0$
induces larger $U>0$, pushing the analyte away. At higher salt
$\Sigma_{L}$, the EOF is mitigated due to the reduction in
$\zeta$-potential (or effective surface charge) indicated by
Eq. \ref{S}. The reduction of EOF to modest values allows the
attraction from the term $q\Phi_{+}$ to overcome the repulsive effect
of the EOF (cf. Fig. \ref{FIG4}(a)), increasing $\theta$. (b)
Occupation fraction $\theta$ as a function of bias voltage $V$ at
different salt ratios and $\Gamma_{R}=0.1$.}
\label{FIG5}
\end{figure}


Since many analyte capture experiments exhibit infrequent pore
blocking, even with bias voltages as high as +250mV ($V \approx +10$),
we use parameters that yield small numerical values of $\theta$.
Henceforth, we set the relative analyte/ion diffusivity $D=0.01$,
$\rho_{\infty} = 10^{-7}$ (corresponding to an analyte concentration
of $\sim 1$nM for $a \sim 5$nm), and the effective analyte charge
$q=30$ (corresponding to that of an approximately 500bp strand of
dsDNA \cite{DNAQ}).  Figure \ref{FIG5} shows representative numerical
solutions of Eq. \ref{TRANS} as a function of (a) salt ratio
$\Sigma_{L}$, and (b) bias voltage $V$.  Fig. \ref{FIG5}(a) shows that
for larger $\Gamma_{R}$ (larger $U$), the repulsive $U/(2D)$ term
dominates in keeping $\theta$ small.  However, upon increasing
$\Sigma_{L}$, attraction arising from an increasing $q\Phi_{+}$ term
eventually increases $\theta$.  Larger values of $\Sigma_{L}$ also
attenuate the pore-averaged $\zeta$-potential, reducing the repulsive
EOF, further increasing $\theta$.  For fixed $V$, we will show that
the analyte capture rate is proportional to $\theta$; therefore,
Fig. \ref{FIG5}(a) predicts the capture rate as a function of salt
ratio.



To determine how $\theta$ depends on the applied voltage, we
use simple assumptions to approximate how the kinetic rates depend on $V$,

\begin{equation}
\koff = \omega_{\rm off}e^{-fqV}, \, \kon = \omega_{\rm on}, \, 
\kt = \omega_{\rm t}V, \nonumber
\end{equation}

\noindent where $\omega_{\rm off}, \omega_{\rm on}$ are intrinsic
detachment and attachment rates, and $1/\omega_{\rm t}$ is the typical
conditional mean time for analyte translocation across the nanopore
under a $v \approx k_{B}T$ voltage bias (translocation is assumed
negligible when $V=0$ \cite{BAYLEY2008} but can be approximated for
polymer translocation \cite{PORET2,PORET3}).  When the macromolecular
analyte blocks the pore and $\theta=1$, a fraction $f$ of the $q$
charges may be exposed to the potential in the pore. When the pore is
completely blocked, this potential is approximately $V$ since there is
no voltage drop across the left reservoir and the pore.  For
detachment to occur, an energy barrier $fqV$ must be overcome,
resulting in $\koff \approx \omega_{\rm off}e^{-fqV}$\cite{SCHULTENF}.
%
%
%
In Fig. \ref{FIG5}(b), $\theta$ is plotted as a function of voltage
$V$ for various fixed salt ratios $\Sigma_{L}$.  The kinetic
parameters chosen were $f=0.02$, $\omega_{\rm off} = \omega_{\rm on}$,
and $\omega_{\rm off}/\omega_{\rm t} = 10^{5}$.  This choice of
intrinsic rates corresponds to approximately 4\% of captured analyte
being translocated at $V=10$, and 96\% detaching back into the bulk
analyte reservoir. For the chosen parameters, increasing a small
voltage $V$ raises $\theta$ exponentially if there is an appreciable
ratio $\Sigma_{L}$ that enhances the positive EOF velocity $U$ to
increase $\Phi_{+}$ beyond the repulsion caused by the positive flow
$U$.  At larger $V$, the larger EOF velocity not only pushes the
analyte faster from the pore, but also contributes to reduction in the
potential $\Phi_{+}$, resulting in a slower increase in 
occupation.

\subsection{Analyte capture rates}

We now define the mean analyte capture rates measured in experiments.
The average times that a pore stays open and blocked are

\begin{equation}
T_{\rm o}\approx {1 \over \kon \rho(1)} = {1-\theta \over
(\koff+\kt)\theta} \quad \mbox{and}\quad T_{\rm b} \approx {1 \over
\koff +\kt},
\label{TIMES}
\end{equation}
respectively. The inverse of the mean time between successive capture
events defines the capture rate $\omega_{\rm c}$:

\begin{equation}
\omega_{\rm c} \approx {1 \over T_{\rm b} + T_{\rm o}}  = (\koff + \kt)\theta.
\label{KCEQN}
\end{equation}
Upon defining the normalized capture rate $\Omega_{\rm c} \equiv
\oc/\omega_{\rm off}$, we formally express

\begin{equation}
\Omega_{\rm c} \equiv \left({\omega_{\rm c}\over \omega_{\rm off}}\right) 
\approx \left(e^{-fqV} + {\omega_{\rm t}\over \omega_{\rm off}}V\right)
\theta,
\end{equation}
where the voltage-dependent expressions for $\koff$ and $\kt$ have
been used and the occupation $\theta$ is determined as a function of
all physical parameters
($\Gamma_{R},f,q,V,w,\Sigma_{L},\rho_{\infty},\omega_{\rm off},
\omega_{\rm on}$, and $\omega_{\rm t}$) by solving Eq. \ref{TRANS}, or
using Eqs. \ref{THETAEQ} or \ref{THETAABSORB}. Thus, for fixed applied
voltage $V$, the capture rate as a function of salt ratio $\Sigma_{L}$
is proportional to $\theta$, and is plotted in Fig. \ref{FIG5}(a).

\begin{figure}[htb]
    \begin{center} 
     \includegraphics[width=3.4in]{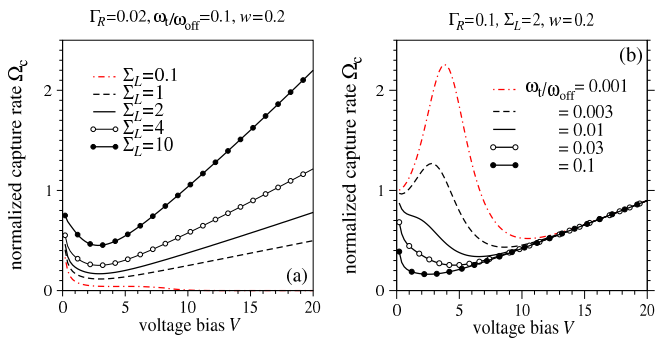}
    \end{center} 
\caption{(a) Normalized capture rate $\Omega_{\rm c}$ as a function of 
bias voltage. (a) The effect of varying salt ratio $\Sigma_{L}$. 
(b) The effect of varying $\kon/\kt = \omega_{\rm on}/\omega_{\rm t}$
with fixed $\Sigma_{L}=2$.}
\label{KC}
\end{figure}

Figure \ref{KC}(a) shows the normalized capture rate $\Omega_{\rm c}$
as a function of voltage $V$ for different salt ratios $\Sigma_{L}$.
For fixed $\Sigma_{L}>1$, increasing $V$ initially decreases
$\Omega_{\rm c}$ by virtue of the prefactor $\omega_{\rm off}
e^{-fqV}$. Only at larger $V$ does the term $\omega_{\rm t}V/\omega_{\rm
off}$ increase the overall capture rate. For small $\Sigma_{L}$, not
only is $\Phi_{+}$ slightly decreased, the $\zeta$-potential and EOF
are increased, repelling the analyte away from the pore, particularly
at high voltages.

Fig. \ref{KC}(b) plots the capture rate as a function of voltage at
various relative translocation rates. For modest $\Gamma_{R}$ and
$\Sigma_{L}$, and relatively small translocation rates $\omega_{\rm
t}/\omega_{\rm off}$, the initial increase in $\Omega_{\rm c}$ with
$V$ arises predominantly from an increase in small $\theta$
(Fig. \ref{FIG5}) which decreases $T_{\rm o}$ (Eq. \ref{TIMES})
despite the decrease in $\koff$.  However, for larger $V$, the
decrease in $\koff$ is not compensated by the slower increase in
$\theta$ (Fig. \ref{FIG5}). Only at large voltages does the $\kt =
\omega_{\rm t}V$ term come into play to increase $\Omega_{\rm c}$
linearly. For larger translocation rates $\omega_{\rm t}$, the pore is
cleared faster by annihilation of the analyte into the opposing
reservoir, preventing the initial increase in $\theta$ with $V$, as
well as the initial decrease in $T_{\rm o}$ that would increase the
overall capture rate.

\section{Summary and Conclusions}

We have modeled the underlying electrokinetics to quantitatively
describe capture of charged analytes by nanopores in the presence of
salt gradients.
%
%
Our analytic analysis shows that the electrostatic potential near the
pore mouth, often neglected, can be sufficient to be an important
determinant in the capture of charged particles.  We also showed that
imposed salt gradients locally change solution conductivity locally,
altering the potential distributions. In particular, higher salt
(higher conductivity) in the non-analyte chamber decreases the voltage
drop in that reservoir and across the nanopore, increasing the analyte
blocking probability.  As a function of salt ratio, our analysis at
small analyte concentrations predicts that both the blocking
probability $\theta$ and the capture rate $\Omega_{\rm c}$ always
increases as $\Sigma_{L}$ increases.  The basic mechanism provides a
physically consistent and testable explanation for recently observed
increases in capture rate with salt ratio $\Sigma_{L}>1$
\cite{MELLERABS}.

Electroosmotic flow also affects analyte capture. By itself,
hydrodynamic flow ({\it e.g.}, from electroosmosis) into the analyte
reservoir sweeps particles away from the pore, dramatically lowering
the blocking probability.  However, when the non-analyte chamber
contains a higher salt concentration such that $\Sigma_{L} > 1$, the
same repelling fluid flow can also change the local conductivity
structure such that the potential $\Phi_{+}$ felt by the charged
analyte at the pore mouth initially increases with flow rate. For
small pore charge/permeability $\Gamma_{R}$, we find that repelling
EOF actually increases the overall {\it attraction} of charged
analytes, particularly when $\Sigma_{L}$ and the effective analyte
charge $q$ are large. Finally, our analysis shows that the capture
rate is sensitive to the translocation rate $\omega_{\rm t}$.  When
$\omega_{\rm t} \gg \omega_{\rm off}$, and nearly all captured
particles are annihilated via translocation into the receiving
reservoir, the capture rate increases with bias voltage $V$, except at
very low $V$.

We considered only outer solutions, accurate in regions where
$Q(\r)\approx 0$ outside the charged boundary layer at the
solution-membrane interface. However, since we focus on the capture of
analyte from the bulk reservoir, the main factor is the voltage drop
across the pore with other electrostatic details within the pore
relatively unimportant.  Although our outer solutions do not hold
inside pores with small radii $a < \kappa_{R}^{-1}$, the ionic current
flow across such pores still induces a slowly decaying electrostatic
potential $\Phi_{r}(r)$ in the bulk analyte chamber, albeit with a
small amplitude determined by an effective, small pore aspect ratio
$w$. The correspondingly smaller potential $\Phi_{r}(r)$ in the right
chamber would give a smaller capture rate at the same analyte
density. We expect the analyte capture rate by smaller pores to have
the same functional dependences as in our mathematical model, but with
a smaller effective $w$.  Note that the EOF velocity $U$ is
proportional to $w^2$ (through $\Gamma(U,\Sigma_{L})$ in Eq. \ref{S}),
while $\Phi_{+}$ scales as $w$. Therefore, we also expect EOF to
become less important than direct electrostatic effects for
sufficiently small pores.

In addition to the charge-neutral approximation, our analysis relies
on a number of other assumptions, including an effective analyte
charge $q$ and pore surface charge $\sigma$ that are independent of
the local ionic strength $\Sigma(\r)$. Moreover, we have assumed
right-cylindrical pores, that the occupation and bulk analyte density
can be approximated using a mean field assumption, and that the
molecular details of the capture and translocation can be described
using simple kinetic rates.  Although some of these assumptions can be
lifted in more detailed models and numerical analyses (for example, in
a model of conical pores \cite{CONICAL}), our simple model embodies
the essential physics of the problem and we expect our results to be
qualitatively predictive.

\vspace{4mm}

\noindent {\bf Acknowledgments} This work was supported by the NSF
through grant DMS-0349195, and by the NIH through grant K25AI41935.


\begin{thebibliography}{99}

\bibitem{BAYLEY2008} Maglia G, Restrepo, MR, Mikhailova E, and Bayley
H (2008) Enhanced translocation of single DNA molecules through
$\alpha$-hemolysin nanopores by manipulation of internal charge. {\it
Proc. Natl Acad. Sci. USA} 105:19720-19725.

\bibitem{GOLOVCHENKO} Gershow M and Golovchenko JA (2007) Recapturing
and trapping single molecules with a solid-state nanopore.  {\it
Nature Nanotechnology} 2:775-779.

\bibitem{MARZIALI} Tabard-Cossa V, Trivedi D, Wiggin M, Jetha NN, and
Marziali A (2007) Noise analysis and reduction in solid-state
nanopores. {\it Nanotechnology} 18:1-6.

\bibitem{DEKKER} Strom AJ {\it et al.} (2005) Fast DNA Translocation
through a Solid-State Nanopore.  {\it Nano Lett.} 5:1193-1197.

\bibitem{MELLERABS} Meller A (2008) Sensing biomolecules translocation
dynamics with solid state nanopores.  PHYS 178, The 236th ACS National
Meeting, Philadelphia, PA, August 17-21, 2008

\bibitem{WANUNUABS} Wanunu M, Cohen-Karni D, Sutin J and Meller A
(2008) Nanopore Analysis of Biopolymers under Physiological Ionic
Strengths.  {\it Biophys. J.} 94:51 (Meeting Abstract)

\bibitem{KLAFTER} Flomenbom O and Klafter J (2003) Single stranded DNA
translocation through a nanopore: A master equation approach.
{\it Phys. Rev.} E 68:041910.

\bibitem{SCHULTEN} Aksimentiev A, Heng JB, Timp G, 
and Schulten K (2004) Microscopic Kinetics of DNA Translocation through
Synthetic Nanopores. {\it  Biophys. J.} 87:2086-2097.

\bibitem{PARK} Park PJ and Sung W (1998) Polymer translocation induced
by adsorption {\it J. Chem. Phys.} 108:3013-3018.

\bibitem{EISENBERG} Singer A, Gillespie D, Norbury J, and Eisenberg RS
(2008) Singular perturbation analysis of the steady state
Poisson-Nernst-Planck system: applications to ion channels. {\it
Eur. J. Appl. Math.}  9:541-560.

\bibitem{EISENBERG1997} Chen, D, Lear J, and Eisenberg RS (1997)
Permeation through an Open channel. Poisson-Nernst-Planck Theory of a
Synthetic Ionic Channel. {\it Biophys. J.} 72:97-116. 

\bibitem{WONGMUTHU} Wong CTA and Muthukumar M (2007) Polymer capture
by electro-osmotic flow of oppositely charged nanopores. {\it
J. Chem. Phys.} 126:164903.

\bibitem{EISENBERG1992} Chen DP, Barcilon V, and
Eisenberg RS (1992) Constant fields and constant gradients in open ionic
 channels. {\it Biophys. J.} 61:1372-1393.

\bibitem{WEINBAUM} Dagan Z, Weinbaum S, and Pfeffer R (1982) An
infinite-series solution for the creeping motion through an orifice of
finite length.  {\it J. Fluid Mech.} 115:505-523.

\bibitem{PARMET} Parmet IL and Saibel E (1965) Axisymmetric creeping
flow from an orifice in a plane wall.  {\it Comm. Pure \& Appl. Math.}
XVIII:17-23.

\bibitem{KELMAN} Kelman RB (1965) Steady-state diffusion through a
finite pore into an infinite reservoir: An exact solution.  {\it
Bull. Mathematical Biophys.} 27:57-65.

\bibitem{EOF} Rice CL and Whitehead R (1965) Electrokinetic Flow in a
Narrow Cylindrical Capillary.  {\it J. Phys. Chem.} 69:4017-4024.

		

\bibitem{IEEE} Ji F, Zuo C, Zhang P, Zhou D (2005) Analysis of
Electroosmotic Flow with Linear Variable Zeta Potential {\it
Proc. 2005 Int. Conf. on MEMS,NANO, and Smart Systems} (ICMENS'05).

\bibitem{IDOL} Anderson JL, Idol WK (1985) Electroosmosis through
pores with nonuniformly charged walls {\it Chem. Eng. Commun.}
38:93-106.

\bibitem{KENNY} Herr AE, Molho JI, Santiago JG, Mungal MG, Kenny TW,
Garguilo MG (2000) Electroosmotic Capillary Flow with Nonuniform Zeta
Potential {\it Anal. Chem.} 72:1053-1057.





\bibitem{SINCHARGE} Whitman PK and Feke DL (1988) Comparison of the
Surface Charge Behavior of Commercial Silicon Nitride and Silicon
Carbide Powders.  {\it J. Am. Ceram. Soc.} 71:1086-1093.

\bibitem{ADSORPTION1} Chou T and D'Orsogna MR (2007) Multistage
adsorption of diffusing macromolecules and viruses.  {\it
J. Chem. Phys.} 127:105101

\bibitem{ADSORPTION2} Diamant H, Ariel G, and Andelman D (2001)
Kinetics of surfactant adsorption: the free energy approach.  {\it
Colloids Surf.} A 183-185:259


\bibitem{DNAQ} Smith SB and Bendich AJ (1990) Electrophoretic charge
density and persistence length of DNA as measured by fluorescence
microscopy.  {\it Biopolymers} 29:1167-1173.

\bibitem{PORET2} Lakatos G, Chou T, Bergersen B, and Patey GN (2005)
First passage times of driven DNA hairpin unzipping.  {\it Physical
Biology} 2:166-174.

\bibitem{PORET3} Grosberg AY, Nechaev S, Tamm M, and Vasilyev O (2006)
How Long Does It Take to Pull an Ideal Polymer into a Small Hole?
{\it Phys. Rev. Lett.} 96:228105



\bibitem{SCHULTENF} Heng JB, {\it et al.} (2005) Beyond the Gene Chip
{\it Bell labs Technical Journal} 10:5-22.

\bibitem{CONICAL} Cervera J, Schiedt B, and Ramirez P (2005) A
poisson/Nernst-Planck model for ionic transport through synthetic
conical nanopores.  {\it Europhys. Lett.} 71:35-41.

\end{thebibliography}
\end{document}